\documentclass[11pt]{article} 

\usepackage{calc}
\usepackage{amsmath,amssymb,amsthm,amscd}
\numberwithin{equation}{section}
\usepackage[bf]{caption}
\usepackage{longtable}
\usepackage{array}
\usepackage{enumerate}
\usepackage{float}
\usepackage{subfig}
\usepackage{multicol}
\usepackage{epsfig}
\usepackage{graphicx}
\usepackage{multirow}
\usepackage{color}
\usepackage{cite}
\usepackage[width=1.09\textwidth]{caption}
\usepackage{tikz}
\usepackage{hyperref}
\usepackage{tikz,textcomp}
\usetikzlibrary{shapes,arrows,positioning,chains,calc}

\usepackage[all,cmtip]{xy}
\usetikzlibrary{matrix, arrows}

\newcommand{\beq}{\begin{equation}}
\newcommand{\eeq}{\end{equation}}
\newcommand{\bea}{\begin{eqnarray}}
\newcommand{\eea}{\end{eqnarray}}

 \usepackage[OT2,T1]{fontenc}
 \DeclareSymbolFont{cyrletters}{OT2}{wncyr}{m}{n}		
 \DeclareMathSymbol{\Sha}{\mathalpha}{cyrletters}{"58}		

\begin{document}

\baselineskip=15pt
\begin{titlepage}

\begin{center}
\vspace*{ 2.0cm}
{\Large {\bf Calabi-Yau Genus-One Fibrations and Twisted Dimensional Reductions of F-theory}}\\[12pt]
\vspace{-0.1cm}
\bigskip
\bigskip 
{
{{Lara B.~Anderson}$^{\,\text{a }}$},  {{James~Gray}$^{\,\text{a }}$} and {{Paul-Konstantin~Oehlmann}$^{\,\text{b }}$}
\bigskip }\\[3pt]
\vspace{0.cm}
{\it 
 ${}^{\text{a}}$ Physics Department,~Robeson Hall,~Virginia Tech,~Blacksburg,~VA 24061,~USA   \\
  ${}^{\text{b}}$  Department of Physics \& Department of Mathematics, Northeastern University \\
360 Huntington Avenue, Boston, MA 02115, United States   
}
\\[2.0cm]
\end{center}

\begin{abstract}
\noindent In this brief note we explore the space of genus one and elliptic fibrations within CY manifolds, their organizing principles, and how they relate to the set of all CY manifolds. We provide examples of genus one fibered manifolds that exhibit different Hodge numbers -- and physically lead to different gauge groups - than their Jacobian fibrations. We suggest a physical mechanism for understanding this difference in twisted circle reductions of $6$-dimensional compactifications of F-theory. \end{abstract}

\end{titlepage}
\clearpage
\setcounter{footnote}{0}
\setcounter{tocdepth}{2}

\section{Introduction}

Much recent work has been devoted to mapping out the space of string vacua, in the context of the string Swampland program and beyond. Playing a central role in this study, string compactifications sweep out a rich landscape of possible solutions. In this context the question of \emph{which effective field theories} can arise is intrinsically linked to the question of \emph{which geometries} can serve as backgrounds for string dimensional reductions? In this work we will explore briefly some of what is understood about one of the simplest classes of such of string geometries -- Calabi-Yau (CY) manifolds and how attempts to bound/characterize Calabi-Yau geometry can shed light on the space of possible M-/F-theory vacua in $6$-/$5$-dimensions.

Since the very first examples of string compactifications \cite{Candelas:1985en}, Calabi-Yau geometries have played an important role. Even in this well-studied context though, there remain many open questions about CY geometries -- including whether the class of CY $n$-folds is finite for $n>2$? It is also unknown whether the possible values of topological invariants (i.e. for a CY 3-fold, independent Hodge numbers $h^{1,1},h^{2,1}$, Chern classes $c_2(X)$, etc) are finite/bounded.

One path towards resolving these questions involves the study of genus-one fibered CY manifolds. In particular, the set of genus one fibered CY 3-folds has been proven to be finite \cite{Gross:1993fd,grassi,filipazzi2021boundedness}, and recent progress \cite{Di_Cerbo_2021,birkar2020boundedness} has given evidence of boundedness for genus one fibered CY $4$- and $5$-folds as well. Moreover, recent analysis of existing datasets of CY $3$- and $4$-folds has demonstrated that the vast majority of all CY manifolds exhibit genus-one fibrations \cite{Anderson:2016cdu,Anderson:2016ler,Anderson:2017aux,Anderson:2018kwv,Anderson:2021oth,Braun:2011ux,Huang:2018esr,Huang:2019pne,Gray:2014fla,Rohsiepe:2005qg}. This apparent genericity of fibration structures provides hope that a classifications of CY genus-one fibrations may lead to tools which could be used to establish the finiteness of the set of \emph{all} CY $n$-folds.

Returning to the question of string effective field theories, elliptic and genus one fibered CY geometries are natural backgrounds for F-/M-theory in $6$- and $5$-dimensions \cite{Vafa:1996xn,Morrison:1996na,Morrison:1996pp}. As a result, more complete characterization of these CY geometries makes it possible to attempt to systematically map out a subspace of the space of M/F-theory vacua. In particular, the close relationship between F-/M-theory under a circle ($S^1$) reduction of the 6-dimensional theory has yielded much information about the possible structure of such theories and inspired many recent classifications of both EFTs and SCFT limits (see e.g. \cite{Heckman:2018jxk} and \cite{Jefferson:2018irk} and reference therein).

In this brief work, we will highlight one aspect of genus-one/elliptic CY geometry that has remained relatively unexplored and the consequences for M/F-theory on elliptically and genus one fibered Calabi-Yau threefolds. In particular, we are interested mathematically in the question of \emph{how different can a genus one fibered Calabi-Yau manifold and its nearest elliptic cousin -- its Jacobian, $J(X)$, be?} Genus one fibered geometries remain much less explicitly explored than elliptically fibered manifolds and the notion of a Jacobian fibration plays a key role in characterizing genus one fibrations. As we will review below, many different genus one fibered manifolds can be related to a single elliptic fibration and the set of manifolds related in this way is called a group of ``CY torsors'' (see e.g. \cite{dolgachev_gross}).  Phrasing the question above more broadly, we are interested in how distinct are the manifolds/elements in the group of CY torsors? A physical application of this question arises in the context of $S^1$ reductions of 6-dimensional F-theory effective theories which include flux along the circle and the possibility of \emph{different non-Abelian gauge groups} in the $6$- and $5$-dimensional theories. This will include an F-/M-theory realization of Dynkin diagrams associated to twisted affine Lie groups and automorphisms that ``fold'' them \cite{kac}. In this context we hope to map out a relatively unexplored corner of this string effective field space. This study appeared in \cite{Anderson:2023wkr} and we provide only a sketch of some of these questions and an illustrative example in the following sections.

\section{Elliptically and Genus-one fibered Calabi-Yau geometry and the basic ingredients of F-theory}

In this section we will briefly review some essential results about elliptic/genus one fibered CY manifolds and their consequences for F-/M-theory. In particular, we shall refer to a \emph{genus one fibered} CY $n$-fold as one that can be described via a surjection $\pi:X \to B_{n-1}$ such that the generic fiber is $\pi^{-1}(b)\simeq \mathbb{E}$ (i.e. a genus one complex curve or $T^2$ as a real manifold) for general points $b \in B_{n-1}$. In general such fibrations do not admit sections, only multi-sections (i.e. $m$-sheeted covers of the base manifold $B_{n-1}$). In the case that the fibration \emph{does} admit a section, the generic fibers are elliptic curves and we refer to the fibration as an \emph{elliptic fibration}. It is this latter class of manifolds that plays a key role in F-theory in $6$- and $4$-dimensions in which the elliptic fibers of a CY $3$ or $4$-fold respectively parameterize the behavior of the axio-dilaton of the (strongly coupled) type IIB theory. 

As described above, recent work has attempted to map out how generic genus-one fibered CY 3-folds/4-folds are within the datasets of known CY manifolds. Overwhelmingly, the available datasets have shown that the presence of fibration structures (i.e. CY $m$-fold fibrations within CY $n$-folds with $m<n$) is ubiquitous for large\footnote{In fact, fibrations seem to occur for all manifolds with $h^{1,1}>19$ in known CY datasets where classifications have been undertaken (see e.g. \cite{Anderson:2017aux}).} $h^{1,1}$. Moreover, a given CY manifold not only appears to be describable as a fibration, generic CY manifolds appear to admit \emph{multiple} such descriptions. In \cite{Gray:2014fla,Anderson:2017aux},  \emph{multiply elliptically fibered} (or genus one fibered in the case without section) CY $n$-folds were described as manifolds admitting multiple descriptions of the form $\pi_i: X_{n} \longrightarrow B^{(i)}_{n-1}$ with elliptic fiber $\mathbb{E}_{(i)b}=\pi^{-1}(b\in B^{(i)}_{n-1})$ (denoted succinctly by $\pi_i: X_{n} \stackrel{\mathbb{E}_{(i)}}{\longrightarrow} B^{(i)}_{n-1}$).  That is,
 \beq\label{manyfibs}
\xymatrix{
& X_{n} \ar[ld]_{\mathbb{E}_{(1)}} \ar[d]^{\mathbb{E}_{(2)}}  \ar[rd]^{\mathbb{E}_{(i)}} &\\
B^{(1)}_{n-1} & B^{(2)}_{n-1} \ldots & B^{(i)}_{n-1}}
\eeq
It is important to note that for each fibration, $\pi_i$, the structure of the singular fibers, discriminant locus, and number of rational sections can all be different, as can the topology of the base manifolds $B^{(i)}_{n-1}$. This can lead to a rich array of consequences for string dualities. Initial steps to explore such prolific fibration structures and their physical consequences were taken in \cite{Gray:2014fla, Anderson:2016cdu,Anderson:2017aux} and the results from \cite{Anderson:2017aux} for one dataset are shown in Figure \ref{ellfib2} below.

\begin{figure}[!h]\centering
\includegraphics[width=0.5\textwidth]{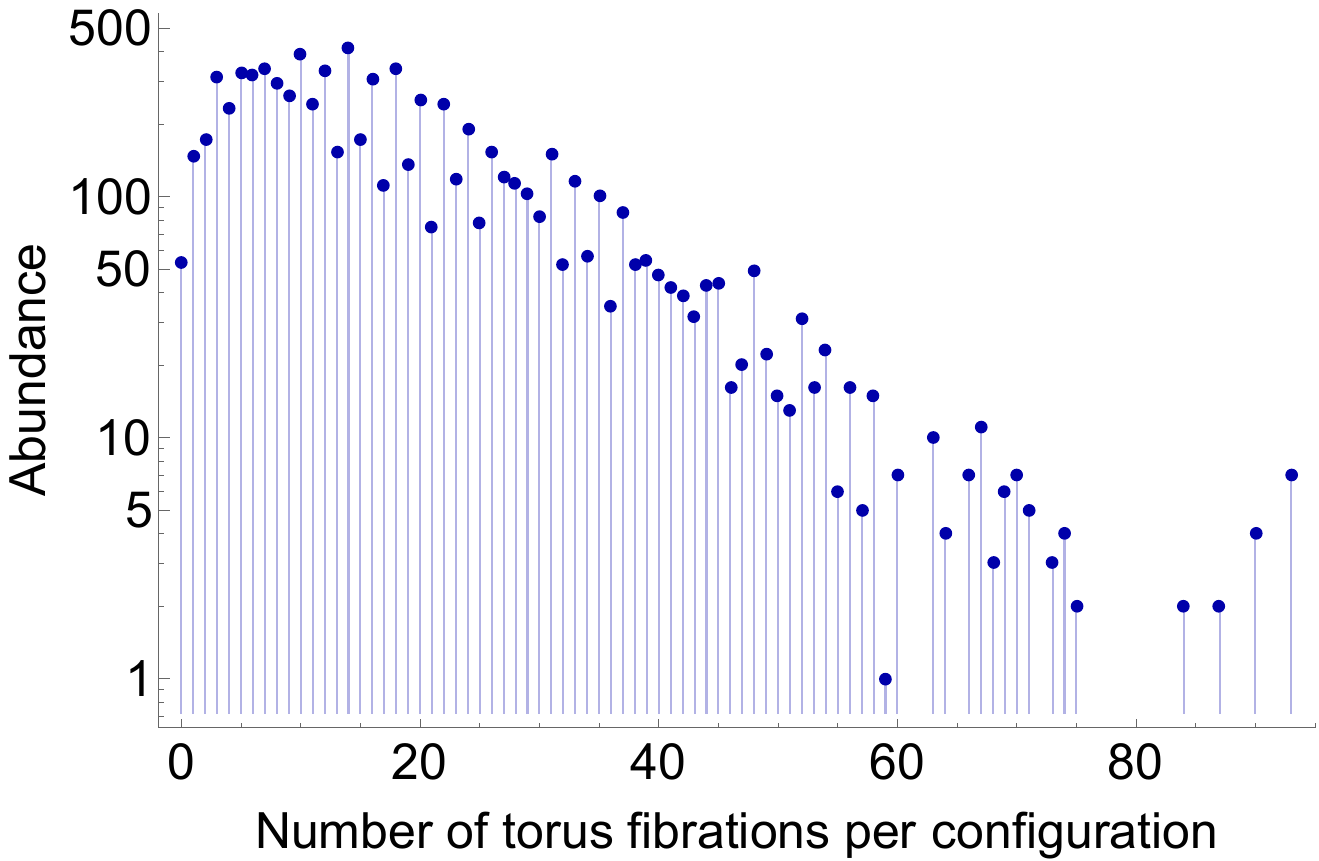}
\caption{{\it Distribution of obvious torus fibrations in the set of CY 3-folds described as complete intersections in products of projective spaces from \cite{Anderson:2017aux}. The values lie in the range 0 - 93. The search in \cite{Anderson:2017aux} yielded $139,597$ fibrations in total.}}
\label{ellfib2}
\end{figure}
Elliptically fibered manifolds exhibit a simple characterization mathematically in that each such manifold is birational to a specified minimal form, a so-called \emph{Weierstrass model} \cite{Nakayama}. Schematically,  a Weierstrass model for an elliptic fibration is built as a hypersurface constraint (with cubic fiber) within a projectivization of three line bundles $\mathbb{P}({\cal O} \oplus {\cal L}^2 \oplus {\cal L}^3)$ with ${\cal L}=K_{B_{n-1}}^{-1}$ in the CY case. This minimal form is achieved by blowing down all components of reducible fibers which do not intersect the zero-section to the elliptic fibration, leaving a generically singular, but simple CY variety. 

Another useful property of elliptic fibrations is that the Picard lattice/K\"ahler forms in such a space can be sorted according to their vertical/horizontal form relative to the elliptic fibration. Namely, each divisor belongs to one of four types \cite{Shioda_Tate_Wazir}: (1) The divisors of the base, $B_{n-1}$ pulled back to $X$, (2) ``fibral" divisors associated to blow-ups of singular elliptic fibers, (3) the zero-section and (4) the free part of the so-called Mordell-Weil group of rational sections (see e.g. \cite{silverman}). Within F-theory in $6$-dimensional theories the number of ``fibral" divisors determine the rank of a non-Abelian gauge group, while the rank of the Mordell-Weil group determines the number of Abelian factors in the effective theory. Elliptic CY fibrations have played a central role in F-theory for the past several decades and many systematic datasets have been constructed \cite{Morrison:2012js,Martini:2014iza,Anderson:2014gla,Halverson:2015jua}.

In contrast, the set of genus-one fibered CY manifolds has remained more obscure until recently. The absence of a zero-section to the fibration makes it difficult to interpret these geometries directly as F-theory backgrounds in any dimension. However, work in recent years has characterized explicitly how they can be viewed as solutions of $M$-theory in one dimension lower. The precise relationship between M-theory on a genus one fibered manifold in $(m-1)$-dimensions and $m$-dimensional solutions of F-theory on its elliptically fibered ``cousin" has been very well studied in the context of Abelian gauge symmetries (see e.g. \cite{Braun:2014oya,Morrison:2014era,Anderson:2014yva,Mayrhofer:2014laa,Cvetic:2015moa,Cvetic:2017epq,Cvetic:2018bni})

From a mathematical point of view, given a genus one fibered CY manifold, $X$, the elliptic CY geometry most closely resembling it is the so-called ``Jacobian fibration'' associated to $X$. The Jacobian fibration, $J(X)$ is the fiber-wise Jacobian of each genus one fiber of $X$. By construction, the Jacobian of a genus-one fibered CY manifold is an elliptically fibered manifold which shares the same discriminant locus to the fibration (and the same j-invariant associated to each fiber). In general, for CY threefolds $J(X)$ is birational to $X$ (and shares the same value for $h^{2,1}$) and furthermore, multiple distinct genus one fibered manifolds can share \emph{the same} Jacobian. The set of such geometries (together with information about the form the birational mapping takes) form a group of ``CY Torsors" \cite{dolgachev_gross}. In this context the Tate-Schaferavich group and the Weil-Chatelet group have played roles in the literature (see e.g. \cite{Bhardwaj:2015oru}) as examples of CY torsors.

While many systematic constructions of datasets of elliptically fibered CY manifolds exist (see e.g. \cite{Morrison:2012js}) no such systematic dataset has been built for genus-one fibered manifolds. One step in this direction would be to understand how to build \emph{all} genus one fibered manifolds in the same CY torsor group -- i.e. given a CY elliptic fibration, $X_0$, is it possible to construct\footnote{It should be noted that the inverse problem -- constructing Jacobians, given an explicit description of the genus one fiber -- is better understood. See e.g. \cite{artin_tate,Braun:2014qka,Knapp:2021vkm}.} all genus one fibered manifolds $X_i$ which have $X_0$ as their Jacobian: $J(X_i)=X_0$? To this end, in the next sections we will explore an explicit example in which a genus one fibered manifold and its Jacobian are quite different indeed and the new physics that this must involve in F-/M-theory. 

\section{$S^1$-dimensional reductions of F-theory in $6$-dimensions and $5$-dimensional M-theory and twisted boundary conditions}\label{sec:boundary_conds}
Let us begin by briefly reviewing a now well-established story involving F-theory circle reductions with \emph{Abelian gauge groups} and M-theory on genus one fibered Calabi-Yau manifolds (see \cite{Cvetic:2018bni} for a review). We will outline a few key ideas here and leave a more detailed discussion for \cite{Anderson:2023wkr}. As initially explored in \cite{Morrison:2014era,Mayrhofer:2014laa}, a 5-dimensional compactification of M-theory on a genus one fibered Calabi-Yau threefold, $X$ is closely related to a 6-dimensional compactification of F-theory on the Jacobian of $X$, the elliptically fibered manifold $J(X)$ described above.

In terms of physics, a compactification of F-theory \emph{requires} the existence of a section in addition to a $T^2$ fiber and thus, we can consider defining F-theory on a given elliptically fibered CY threefold. However, this single compactification of F-theory can give rise to a number of distinct 5-dimensional compactifications of M-theory via circle reduction with flux (and possibly matter field vevs) \cite{Witten:1996qb,Grimm:2015zea,}. In particular, if the $6$-dimensional effective theory arising from F-theory is Abelian, then it is possible to consider a circle flux (i.e. discrete holonomy), 
$\xi=\int_{S^1} A$. Different choices of $\xi$ lead to different compactifications of M-theory on (possibly) distinct genus-one manifolds which are different elements of the group of CY torsors\footnote{In simple cases the Tate-Shaferavich or Weil-Chatelet groups.} (note that the reduction with $\xi=0$ leads to an M-theory compactification on the elliptically fibered manifold $J(X)$ itself)  \cite{Morrison:2014era,Anderson:2014yva,Cvetic:2015moa}. The set of these 5-dimensional theories are connected by geometric transitions in the underlying genus one fibered manifolds. This framework has been used to gain a deep understanding of discrete symmetries in F-/M-theory\footnote{See \cite{Mayrhofer:2014laa} for links to discrete torsion in genus-one/elliptically fibered CY geometries, \cite{Schimannek:2021pau} to non-commutative resolutions and \cite{Dierigl:2022zll} for its relation to discrete gauge anomalies.
}.

In the context of the questions raised here about Jacobian fibrations and groups of CY torsors, we are interested in manifolds that also contain fibral divisors. That is, compactifications of F-theory in 6-dimensions which include non-Abelian gauge group factors, in addition to the possible existence of multiple rational sections, i.e. symmetry groups of the form $G\times U(1)$ (possibly with discrete group quotients or multiple Abelian factors as well). In this case, the circle reductions can involve flux from gauge fields in the Cartan subgroup of $G$ as well as the $U(1)$ factors. And as we will see, a new feature that is possible in this context is the presence of boundary conditions which can mix these different fluxes together in non-trivial ways. As an action on gauge fields this is schematically of the form
\beq\label{sigma}
A^B(s)=\sigma(A^C(s+2\pi))
\eeq
\noindent where $\sigma$ is a discrete, outer automorphism of $G \times U(1)$. These boundary conditions can reduce the $5$-dimensional symmetry down to a subgroup.

Such non-trivial boundary conditions (and fluxes) have been studied in a variety of other contexts under the name of ``twisted'' circle reductions (see e.g. \cite{Tachikawa:2011ch,Kim:2019dqn,Kim:2004xx,Bhardwaj:2019fzv,Lee:2022uiq}) but has not been previously implemented in the compact setting in F-theory. We find that in the presence of these boundary conditions we generate a new set of 5-dimensional vacua associated to M-theory compactifications on genus-one fibered manifolds with different non-Abelian gauge groups of the form $H_i \times U(1)$ (or $H_i \times \mathbb{Z}_n$) where $H_i \subseteq G$ is a sub-group. Geometrically, these geometries are novel in that the genus one fibered manifold and its Jacobian can realize \emph{different} non-Abelian gauge groups, despite the fact that they share a discriminant locus (see also \cite{Braun:2014oya,Anderson:2019kmx,Oehlmann:2019ohh,Baume:2017hxm,Kohl:2021rxy}).

In \cite{Anderson:2023wkr} we find that the possible groups $H,G$ that can arise are highly constrained and the discrete boundary conditions allowed closely relate to ``foldings'' of the Dynkin diagrams (see Figure \ref{fig:A2Folding} below). Moreover, in the context of the genus one fibered geometries considered here we find that the discrete automorphisms in play are in fact not simply those of the Dynkin diagram associated to the Lie group G itself, rather we find geometric realizations of the folding of the Dynkin diagrams associated to affine groups $\tilde{G}$). This is perhaps to be expected since it is the affine Dynkin diagrams which are realized as components of the resolved elliptic fibrations. To illustrate this conceretely, we turn now to a simple example.

\section{An Illustrative Example}
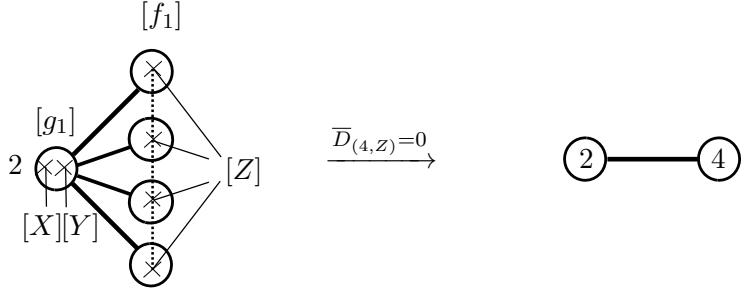
\begin{figure}[t!]
 \begin{center}
 \begin{picture}(00, 110)
 \put(-150,10){\includegraphics[scale=0.6]{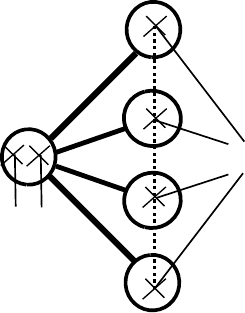} }
 \put(-150,70){$[g_1]$}
  \put(-160,53){$2$}
  \put(-155,30){$[X]$}
  	  \put(-140,30){$[Y]$}
 \put(-110,110){$[f_1]$}
 \put(-78,51){$[Z]$} 
 \put(-40,55){$\xrightarrow{\overline{D}_{(4,Z)}=0}$     }
 \put(50,50){\includegraphics[scale=0.6]{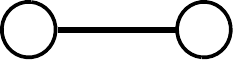} }
 
  \put(56,55){$2$} 
    \put(106,55){$4$} 
 \end{picture} 
 \caption{{\it \label{fig:A2Folding}Depiction of an $A_2^{(2)}$ fiber. The two two-sections $X, Y$ intersect the  middle node. All four outer roots are identified by the four-section $Z$. At the intersection with the monodromy divisor the fiber degenerates to a multiplicity two fiber.}}
 \end{center}
 \end{figure}

In this section we illustrate a genus one geometry that exhibits one of the simplest twisted affine fibrations possible. This CY 3-fold is a genus one fibration $\pi: X \to \mathbb{P}^2$ and is described by the following reflexive polytope \begin{align}
\begin{array}{cc}
\begin{array}{|c|c|}\hline
X&(-1,1,0,0) \\ 
Y&(-1,-1,0,0)\\ 
Z&(1,0,0,0)\\ \hline
\end{array}
&
\begin{array}{|c|c|}\hline
f_1&(1,0,1,0)\\ 
g_1&(1,0,2,0)\\ 
z_1&(-3,0,-1,-1)\\ 
z_2&(0,0,0,1)\\ \hline
\end{array}
\end{array}
\end{align}
which allows the computation of the Hodge numbers via the Batyrev formula as
$(h^{1,1},h^{2,1}) = (3,107)$.
The genus-one fiber is of quartic type and admits an affine $A_2^{(2)}$ fiber whose associated finite group is $SU(2)$ given as
{\small \begin{align}
p=d_1   X^4 + d_2   X^3 Y + d_3   X^2 Y^2 + d_4   X Y^3 + d_5   Y^4 + 
 d_6   f_1   g_1 X^2 Z + d_7 f_1   g_1 X Y Z + d_8   f_1   g_1 Y^2 Z + f_1 Z^2
\end{align}}
where ${f_1,g_1}$ are the fibral divisors that resolve the $A_2^{(2)}$ type of fiber and the polynomials $d_i$ are valued in the $\mathbb{P}^2_{z_0,z_1,z_2}$. The fibral divisors project on $z_0$ as
\begin{align}
\{ f_1, g_2  \} \xrightarrow{\pi} z_0=f_1 g_1^2 \, . 
\end{align} 
 The Stanley-Reisner ideal is determined by
\begin{align}
\mathcal{SRI}: \{ Z g_1, Y X Z, Y X f_1, g_1 z_1 z_2, f_1 z_1 z_2  \}\, .
\end{align}
Here we note, that the divisor $f_1=0$ contains four components that are intersected by the four-section $Z=0$ given as the discriminant locus
\begin{align}
p_{|Z=0}=&d_1   X^4 + d_2 X^3 Y + d_3 X^2 Y^2 + d_4 X Y^3 + d_5 Y^4 \, , \\
p_{|f_1=0}=&d_1 X^4 + d_2 X^3 Y + d_3 X^2 Y^2 + d_4 X Y^3 + d_5 Y^4 \, .
\end{align}
These four-components are interchanged along the discriminant locus $Disc(p_{|Z=0})$. The components in general intersect as a $A_2^{(2)}$ form fiber, and notably over $f_1 = Disc(p_{ |Z=0})=0$, the four components degenerate to a single $\mathbb{P}^1$ and hence the fiber admits the multiplicities $(4,2)$ and therefore a multiple fiber \cite{gross3foldsII,Anderson:2018heq}. The monodromy which reduces the symmetry to $A_{2}^{(2)}$ is controlled by the following locus in the base:
\begin{align}
\begin{split}
\overline{D}_{4,Z}=& 2 d_ 2^3 (2 d_ 4^3 - 9 d_ 3 d_ 4 d_ 5) + 
     2 d_ 1 d_ 2 d_ 4 (-9 d_ 3 d_ 4^2 + 40 d_ 3^2 d_ 5 + 
        96 d_ 1 d_ 5^2) \\ & + 27 d_ 2^4 d_ 5^2 + 
   d_ 2^2 (-d_ 3^2 d_ 4^2 + 4 d_ 3^3 d_ 5 + 6 d_ 1 d_ 4^2 d_ 5 - 
      144 d_ 1 d_ 3 d_ 5^2) \\ & + 
 d_ 1 (4 d_ 3^3 d_ 4^2 + 27 d_ 1 d_ 4^4 - 
     16 d_ 3 (d_ 3^3 + 9 d_ 1 d_ 4^2) d_ 5 + 128 d_ 1 d_ 3^2 d_ 5^2 - 
     256 d_ 1^2 d_ 5^3)\, .
     \end{split}
\end{align}
See Figure \ref{fig:A2Folding} for an illustration of the ``folding" of this affine Dynkin diagram realized in the fiber. From this perspective, the three K\"ahler moduli are interpreted as the the two classes of fibre and base $\mathbb{P}^2$ as well as the $SU(2)$ gauge algebra. 
 
 Notably, such a four-section does not exist in the generic Weierstrass or Tate model. The respective Jacobian can be obtained by shrinking the $g_1$ curve and using the Jacobian map given below. Indeed, upon shrinking the $g_1$ component to a singular genus-one fibration and taking the Jacobian mapping to produce an elliptic fibration in Weierstrass form \cite{artin_tate} one finds the Weierstrass coefficients
 
{\scriptsize \begin{align}
\begin{split}
f=& \frac{1}{48} f_1^2 (-(4 d_3 - d_7^2 f_1 + 4 d_6 d_8 f_1)^2 - 
   24 (-2 d_2 d_4 + 8 d_1 d_5 - 2 d_5 d_6^2 f_1 + d_4 d_6 d_7 f_1 + 
      d_8 (-2 d_3 d_6 + d_2 d_7 - 2 d_1 d_8) f_1)) \, . \\
g=&\frac{1}{864} f_1^3 (36 (-2 d_2 d_4 + 8 d_1 d_5 - 2 d_5 d_6^2 f_1 + d_4 d_6 d_7 f_1 + 
      d_8 (-2 d_3 d_6 + d_2 d_7 - 2 d_1 d_8) f_1) \times \nonumber \\ &  \qquad \qquad (-4 d_3 + (d_7^2 - 
         4 d_6 d_8) f_1) + (-4 d_3 + (d_7^2 - 4 d_6 d_8) f_1)^3 + \nonumber \\ &
   216 ((d_4^2 - 4 d_3 d_5) d_6^2 f_1 + 2 d_2 d_6 (2 d_5 d_7 - d_4 d_8) f_1 + 
      d_2^2 (-4 d_5 + d_8^2 f_1) - 
      4 d_1 (d_4^2 - 4 d_3 d_5 + d_5 d_7^2 f_1 - d_4 d_7 d_8 f_1 + d_3 d_8^2 f_1)))\, . 
      \end{split}
 \end{align}}
 from which we obtain the discriminant to leading orders as 
 \begin{align}
 \Delta = 4f^3 + 27g^2 =f_1^6 \overline{D}_{4,Z} + \mathcal{O}(f_1^7) \, .
 \end{align}

Here we find the fiber to be of $I_4^*$ non-split form which corresponds to a $G_2$ gauge group in $6$-dimensions \cite{Bershadsky:1996nh,Katz:2011qp}. Such a fiber can be resolved via two more resolution divisors \cite{Esole:2017rgz} and hence we expect the Jacobian two have a four dimensional K\"ahler moduli space. Thus this example illustrates the differences described above between CY torsors in that the $G_2$ gauge symmetry realized in the Jacobian is larger than the $SU(2)$ symmetry found in the genus one fibered manifold. 

\section{Summary}
In this work we have given a brief illustration of the important role played by elliptic/genus-one fibrations both in mathematically characterizing Calabi-Yau manifolds as well as mapping out spaces of string solutions in various dimensions. In particular, we have provided explicit examples of genus one fibered manifolds exhibiting different singular fibers -- and hence corresponding to different physical gauge groups -- compared to their elliptic Jacobian counterparts. In this setting we expect these genus-one fibered geometries to correspond to the twisted dimensional reductions of F-theory described in Section \ref{sec:boundary_conds}. However, there remain a number of interesting questions to be addressed in this context, including 1) How is an outer automorphism of the form shown (\ref{sigma}) linked to the apparent folding of the \emph{affine} Dynkin diagrams which are realized in the elliptic fibers? 2) How are the necessary discrete symmetries realized explicitly in the CY geometry? 3) How are the CY torsors appearing in this setting linked by geometric transitions (i.e. Higgsing transitions in the 5D theory)? We address these and other questions in   \cite{Anderson:2023wkr}. For now we hope that the brief discussion above illustrates interesting open questions -- both physical and mathematical -- remaining within the study of CY fibrations.

\section*{Acknowledgements}
We thank Luca Cassia, Markus Dierigl, Jonathan Mboyo Esole, Antonella Grassi, Jie Gu, Sheldon Katz, David Morrison, Nikhil Raghuram and Thorsten Schimannek, for illuminating discussions. The work of L.A. and J.G. is supported by the NSF grant PHY-2014086. The work of P.K.O.\ is supported by a Swedish grant of the Carl Trygger Foundation for Scientific Research and the European Research Council (ERC) under the European Union's Horizon 2020 research and innovation program (grant agreement No. 851931) and funding from  the NSF CAREER grant PHY-1848089 and startup funding from Northeastern University by Fabian Ruehle.

The authors would like to thank the organizers of the Nankai Symposium on Mathematical Dialogues where the talk inspiring this brief survey was presented.

\end{document}